\documentclass[aps,prl,twocolumn,showpacs,amssymb,amsmath]{revtex4}
\usepackage{graphicx}
\usepackage{epsfig}
\usepackage{bm}

\begin{document}

\title{Dynamic transition in driven vortices across the peak effect in superconductors}
\author{Mahesh Chandran} 
\email{chandran@physics.ucdavis.edu} 
\author{R. T. Scalettar} 
\author{G. T. Zim\'{a}nyi} 
\affiliation{Department of Physics,
University of California, Davis, California 95616 }

\date{\today}

\begin{abstract} 
We study the zero-temperature dynamic transition from the disordered flow to an 
ordered flow state in driven vortices in type-II superconductors. The transition 
current $I_{p}$ is marked by a sharp kink in the $V(I)$ characteristic with a 
concomitant large increase in the defect concentration. On increasing magnetic field 
$B$, the $I_{p}(B)$ follows the behaviour of the critical current $I_{c}(B)$.
Specifically, in the peak effect regime $I_{p}(B)$ increases rapidly along 
with $I_{c}$. We also discuss the effect of varying disorder strength on $I_{p}$.
\end{abstract}

\pacs{74.25.Qt, 74.25.Sv}

\maketitle

The model of driven interacting particles in presence of quenched disorder captures essential
features of the dynamics occurring in many condensed matter system. Of specific interest is the
dynamics of vortices above the critical (or depinning) current $I_c$ in type II superconductors.
Early numerical simulation of the transport characteristics in 2D showed that the topological
defects are generated close to $I_c$, and are annealed at high driving force~\cite{shi}. This
suggested a dynamic transition from the disordered flow to an ordered flow at a current
$I_{p}\!>\!I_{c}$. Theoretically, Koshelev and Vinokur (KV) considered this transition at $I_p$
similar to an equilibrium ``melting'' transition of a clean system in the moving reference
frame~\cite{kosh}. The effect of quenched disorder on the moving vortices is included in the theory
through an effective shaking temperature $T_{sh}\!\propto\!I^{-1}$. In Koshelev-Vinokur theory, the
re-crystallization current $I_p$ increases as $(T_{m}-T)^{-1}$ on approaching the equilibrium
melting temperature $T_m$ of the static system. Later calculations~\cite{scheidl,balents,giamarchi}
and simulations~\cite{moon,olson,kolton,fangohr} identified the free flowing state above $I_p$ as
the smectic phase. In the smectic phase, the particles move in static channels with quasi-long range
order perpendicular to the flow compared with liquid like short range order within the
channels~\cite{comment1}.

The dynamic transition in driven vortices was first studied experimentally in the transport
measurements by Bhattacharya and Higgins (BH)~\cite{shobo}, and was later observed in neutron
scattering~\cite{yaron} and Bitter decoration~\cite{duarte} experiments. BH studied the behaviour
of transition current $I_p$ as a function of the magnetic field $B$, particularly in the peak
effect (PE) regime. The PE is marked by a rapid increase in $I_{c}(B)$, generally close to the
upper critical field $B_{c2}$. BH found that the $I_{p}(B)$ increases rapidly in the field range
in which the PE occurs. This behaviour of $I_{p}(B)$ in the PE regime is similar to the behaviour
of $I_{p}(T)$ around $T_m$ as observed in simulation by KV in Ref.~\cite{kosh}. The similarity can
be traced to the softening of the shear modulus $c_{66}$ of the vortex lattice. But an important
difference separates the two :  unlike BH, KV do not see any corresponding increase in $I_{c}(T)$,
{\it i.e.}, no peak effect in the $I_{c}(T)$ is observed in the simulation. This makes it
difficult to correlate the increase in $I_{p}(T)$ close to $T_m$ to the enhanced coupling of the
vortex lattice to the quenched disorder. The increase in $I_{p}(T)$ could very well be due to
increased thermal fluctuations on approaching $T_m$, thus requiring larger currents to anneal the
lattice defects.

Recently, we showed that the PE occurs in a system of 2D vortices close to $B_{c2}$
at zero temperature~\cite{chandran1}. In this paper, we revisit the dynamic 
transition, particularly across the PE. We find that $I_{p}(B)$ increases rapidly
in the field range in which the $I_{c}(B)$ shows the PE. The dynamic transition at
$I_{p}$ is characterized by hysteresis in $V(I)$ and a sharp peak in the dynamic
resistance $R_{d}(I)$. The topological defect concentration also shows a jump at
$I_p$. We present the dynamic phase $I$-$B$ diagram, and discuss the effect of 
increasing the disorder strength $\Delta$ on $I_{p}$.

Consider a 2D cross-section of a bulk type-II superconductor perpendicular to the 
magnetic field ${\bf B}=B\hat{\bf z}$. Within London's approximation, the 
dynamics of vortices are governed by the overdamped equation of motion 
\begin{eqnarray}
\eta\frac{d{\bf r}_{i}}{dt} = -\sum_{j\neq i} \nabla U^{v}({\bf r}_{i}-{\bf r}_{j})
- \sum_{k} \nabla U^{p}({\bf r}_{i}-{\bf R}_{k}) + {\bf F}_{ext}. \nonumber
\end{eqnarray} 
Here $\eta$ is the flux-flow viscosity, and the first term represents the
inter-vortex interaction given by the potential
$U^{v}(r)=\frac{\phi_{0}^{2}}{8\pi^{2}\lambda^{2}} K_{0}(\tilde{r}/\lambda)$, where
$K_0$ is the zeroth-order Bessel function, and $\tilde{r} =
(r^{2}+2\xi^{2})^{1/2}$. $\lambda$ and $\xi$ are the penetration depth
and coherence length of the superconductor, respectively. $\phi_{0}$ represents the
flux quantum. The vortex pinning is added through the second term which is an 
attractive interaction with parabolic potential wells 
$U^{p}(r)=U_{0}(\frac{r^{2}}{r_{p}^{2}}-1)$ for $r <
r_{p}$, and 0 otherwise, centered at the random ${\bf R}_{k}$ locations. The third
term ${\bf F}_{ext}=\frac{1}{c}{\bf J}\times \phi_{0}\hat{\bf z}$ is the Lorentz
force due to transport current density ${\bf J}$. The length is in units of
$\lambda(B\!=\!0)\!=\!\lambda_{0}$, and $J$ is in units of $cf_{0}/\phi_{0}$
where $f_{0}\!=\!\!\frac{\phi_{0}^{2}}{8\pi^{2}\lambda_{0}^{3}}$. The time $t$ is in 
units of $\eta\lambda/f_{0}$ whereas the velocity $v$ of vortices is in units of 
$f_{0}/\eta$.

We use the reduced magnetic field $b\!=\!B/B_{c2}$ with
$B_{c2}\!\!=\!\!\frac{\phi_{0}}{2\pi\xi^{2}}$, and is calculated from the lattice 
constant
$\frac{a_{0}}{\lambda}\!=\!(\frac{4\pi}{\sqrt{3}})^{\frac{1}{2}}(\frac{1}{\kappa^{2}b})^{\frac{1}{2}}$.
The $B$ dependence of $\lambda$ is follows 
$\lambda(b)=\lambda_{0}/(1-b^{2})^{\frac{1}{2}}$, with a similar expression for $\xi$.
The simulation is for the Ginzburg-Landau parameter
$\kappa\!=\!\frac{\lambda}{\xi}\!=\!10$ which is typical of the low-$T_c$ materials. The
number of vortices $N_v$ was chosen in the range of 800-1200. The prefactor $U_{0}$ of 
the pinning potential is distributed randomly between $\Delta\!\pm\!0.01$. The results 
presented below are for the pin density $n_{p}\!=\!2.315$. We calculate the transport 
characteristics $V(I)$, where the current $I\!\propto\!J_{y}$, and the voltage 
$V\!\propto\!\langle v_{x}\rangle$.  The topological defect concentration $f_{d}(I)$ is 
obtained by Delaunay triangulation of the real space configuration. We choose the free 
flowing vortex lattice at high driving currents as the initial configuration, and 
decrease the current to 0 in small steps across $I_c$. This particular method of 
preparing the system minimizes the influence of non-equilibrium defects which are present 
in field cooled simulations. The equation of motion is time integrated by standard 
techniques. The $V(I)$ and $f_{d}(I)$ are time averaged in the steady state. Parallel 
algorithms were employed to speed up the simulation at high densities~\cite{chandran2}.

\begin{figure}[hbt]
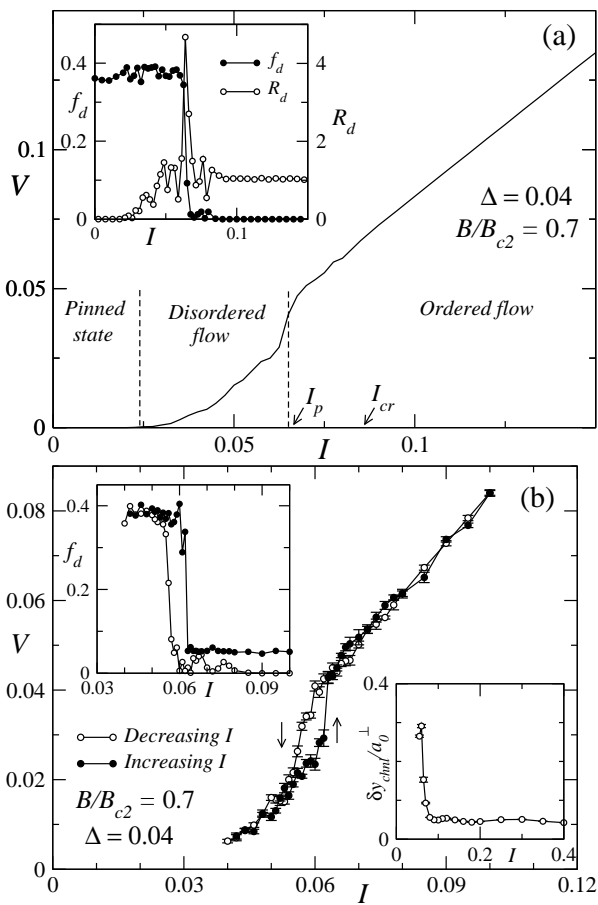
 
\includegraphics[width=225pt]{fig1a} 
\includegraphics[width=225pt]{fig1b} 
\caption{(a) The $V(I)$ characteristics for $\Delta\!=\!0.04$ and $b\!=\!0.7$.
Inset : The defect concentration $f_{d}(I)$, and the dynamic resistance
$R_{d}(I)$. (b) Hysteresis in $V(I)$ and $f_{d}(I)$ (upper inset) across $I_{p}$.
The lower inset shows the $(\delta y_{chnl})/a_{0}^{\perp}$, where
$a_{0}^{\perp}\!=\!\frac{\sqrt 3}{2}a_{0}$ and $a_0$ is the lattice constant.}
\end{figure}

Fig.~1(a) shows the $V(I)$ curve for $b\!=\!0.7$ and $\Delta\!=\!0.04$. Broadly,
three current regimes can be identified:  (a) a pinned state for $I\!<\!I_{c}$,
(b) a disordered flow state between $I_{c}\!<\!I\!<\!I_{p}$ where some vortices
remain immobile and large transverse excursions of the active channels are
present, and (c)  an ordered flow state for $I\!>\!I_{p}$ with all vortices moving
and interchannel hopping takes place with, at most, neighboring channels. The
$f_{d}(I)$, and the dynamic resistance $R_{d}(I)\!=\!\frac{dV}{dI}$ is plotted in
the inset of Fig.~1(a). The current $I_p$ is marked by a sharp peak in $R_{d}(I)$
and a large change in $f_{d}$. The response function $V(I)$ shows a kink at
$I_{p}$, as evident from Fig.~1(a). We find that $V(I)$ and $f_{d}(I)$ are also
hysteretic across $I_{p}$, as shown in Fig.~1(b). The kink and the hysteresis in
the response function $V(I)$ suggests that the dynamic transition occurs at
$I_{p}$. The transition is of first order nature, as was identified in
Ref.~\cite{kosh}.

In Fig.~2, real space configurations are shown for currents above and below
$I_{p}$ corresponding to parameters in Fig.~1. The instantaneous configuration
shows a relatively ordered lattice above $I_{p}$. The vortex trajectories are well
separated channels aligned parallel to $v_{x}$. The average transverse wandering
of the channel $\langle\delta y_{chnl}\rangle < \alpha a_{0}^{\perp}$, where
$a_{0}^{\perp}\!=\!\frac{\sqrt 3}{2}a_{0}$ and $a_0$ is the lattice constant, and
$\alpha\!\approx\!0.05$. Just below $I_{p}$, $\alpha$ becomes $\approx 0.3$, as
shown in the inset of Fig.~1(b). The slowing down of vortices effectively couples
the transverse velocity component $v_{y}$ to the quenched disorder. This induces
large scale dislocations, and the dynamic friction ($\propto R_{d}$) increases.
The increase in transverse wandering of vortices is also reflected in the dense 
braiding of the active channels. The behaviour of $\langle\delta
y_{chnl}\rangle$ across $I_{p}$ is similar to the Lindemann criterion for thermal
melting. Note that some of the vortices appear as immobile for $I\!=\!0.055$ in
Fig.~2(b) (rightside panel). The region of immobile vortices grow with decreasing
$I$ until $I_c$ is reached when the whole system is pinned.

\begin{figure}[hbt] 
\includegraphics[width=210pt]{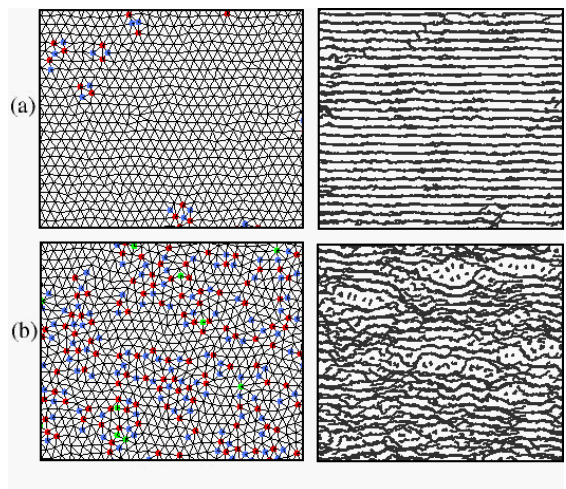} 
\caption{Instantaneous vortex configuration (left) and vortex trajectories (right) for
(a) $I\!=\!0.06\!>\!I_{p}$, and (b) $I\!=\!0.055\!<\!I_{p}$. The $b\!=\!0.7$, and
$\Delta\!=\!0.04$.}
\end{figure}

The $R_{d}(I)$ has been used previously to identify the phase boundary. In
Ref.~\cite{shobo}, the transition from the disordered flow to the ordered flow
state is identified as the current at which $R_{d}(I)\!\approx\!R_{BS}$ where
$R_{BS}$ is the asymptotic Bardeen-Stephen flux-flow resistance. This current is
marked as $I_{cr}$ in Fig.~1(a).  In Ref.\cite{kolton}, the state between $I_{p}$
and $I_{cr}$ is identified as the smectic phase, whereas above $I_{cr}$, the state
is defined as a transversely frozen phase. The later state is distinguished from
the smectic phase by the absence of transverse jumps by vortices between adjacent
channels.

We note that neither $V(I)$ nor $f_{d}(I)$ shows any feature at $I_{cr}$.  
Therefore, a more appropriate description of the state between $I_{p}$ and
$I_{cr}$ is possible in terms of fluctuation which allows occasional transverse
excursions of the vortices between adjacent channels. The rate of such transverse
jumps can be calculated by employing Arrhenius relation which is conventionally
used to describe thermal activation over energy barrier, but with temperature
replaced by $T_{sh}$~\cite{chandran3}. Such a description can account for the
suppression of transverse jumps above $I_{cr}$, and also for the experimentally
observed $R_{d}(I)$ curves. To summarize, the dynamic transition in the moving
lattice occurs at $I_p$ at which the dynamic resistance $R_{d}(I)$ shows a peak.
The flowing state above $I_p$ is a smectic phase with small longitudinal
correlation and quasi-long range order in the transverse direction.

\begin{figure}[hbt] 
\includegraphics[width=210pt]{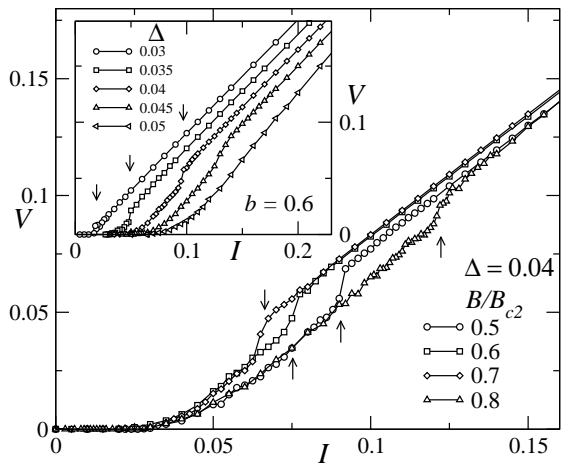}
\caption{$V(I)$ curves for $\Delta\!=\!0.04$ for fields across the PE. Inset :
$V(I)$ curves for increasing disorder strength $\Delta$ for $b\!=\!0.6$. The curves
in the inset are shifted horizontally by 0.01. The arrows mark the transition
current $I_p$.}
\end{figure}

We now consider the behaviour of $I_p$ as the magnetic field $b\!=\!B/B_{c2}$ is
changed, particularly across the PE in $I_{c}(b)$. For $\Delta\!=\!0.04$, the onset
field for PE is $b_{op}\!\approx\!0.75$ whereas the peak occurs at
$b_{peak}\!\approx\!0.9$. Fig.~3 shows the $V(I)$ curves for four values of $b$
between 0.5 and 0.8. The $I_p$ can be easily identified by the sharp kink in the
$V(I)$ curves. With increasing $b$, the $I_p$ first decreases before increasing
rapidly in the PE field regime. The $f_{d}(I)$ and $R_{d}(I)$ characteristics
across $I_p$ are similar to that shown in Fig.~1 for all values of $b$. In the
inset of Fig.~3, the effect of increasing disorder strength $\Delta$ on $I_p$ is
shown for $b\!=\!0.6$. Other than the expected increase in $I_p$ with increasing
$\Delta$, we find that the transition at $I_p$ is broadened for
$\Delta\!\geq\!0.05$. Also, the hysteresis in $V(I)$ across $I_p$ is too small to be
distinguished. This suggests that the first-order nature of the dynamic transition at 
$I_p$ does not survive for $\Delta\!\geq\!0.05$.

Fig.~4(a) shows the dynamic phase diagram $I$-$B$ for $\Delta\!=\!0.04$ which summarizes the
main result of the paper. The $I_{p}(b)$ and $I_{c}(b)$ shows similar behaviour as $b$
approaches the upper critical field value. Particularly, the rapid increase in $I_{p}(b)$
coincides with the peak effect in $I_{c}(b)$. The region between $I_{p}(b)$ and $I_{c}(b)$
constitutes the disordered flow regime, whereas above $I_{p}$ vortices flow in ordered
channels. For $\Delta\!=\!0.04$, the $I_{p}(b)\!=\!k(b_{0}-b)^{-1}$ with $b_{0}\!=\!0.92$ and
$k\!=\!0.015$ gives a reasonable fit for $b\!>\!b_{op}$, as shown by the thick line in
Fig.~4(a). This form of $I_{p}(b)$ was motivated by the behaviour of $I_{p}(T)$ on approaching
the equilibrium melting transition in Ref.~\cite{kosh}. The $I$-$B$ plot in Fig.~4(a)  thus
supports the picture that both {\em the static and the dynamic friction of the vortex system
increases rapidly in the field range in which the peak effect occurs}. As shown in
Ref.~\cite{chandran1}, the PE in the critical current $I_c$ is driven by the softening of the
vortex interaction on approaching the upper critical field $B_{c2}$. Softening of the vortex
interaction also reduces the shaking temperature $T_{sh}\!\propto\!I^{-1}$ required for the
dynamic ordering (or recrystallization) of the vortices which explains the increase in
$I_{p}(b)$ in the PE regime. Thus, an increase in $I_c$ (as a function of $B$ or $T$) implies
an increase in $I_p$. We emphasize that the reverse does not hold, {\it i.e.}, increase in
$I_p$ {\em does not} imply an increase in $I_c$. This can be seen from the behaviour of
$I_{p}(T)$ and $I_{c}(T)$ in Ref.~\cite{kosh} where thermal depinning causes $I_{c}(T)$ to 
decrease monotonically even as $I_{p}(T)$ increases on approaching the equilibrium melting 
temperature $T_m$. Overall, we find the behaviour of $I_{p}(B)$ and $I_{c}(B)$ in good 
agreement with the experimental observation~\cite{shobo}.

\begin{figure}[hbt]
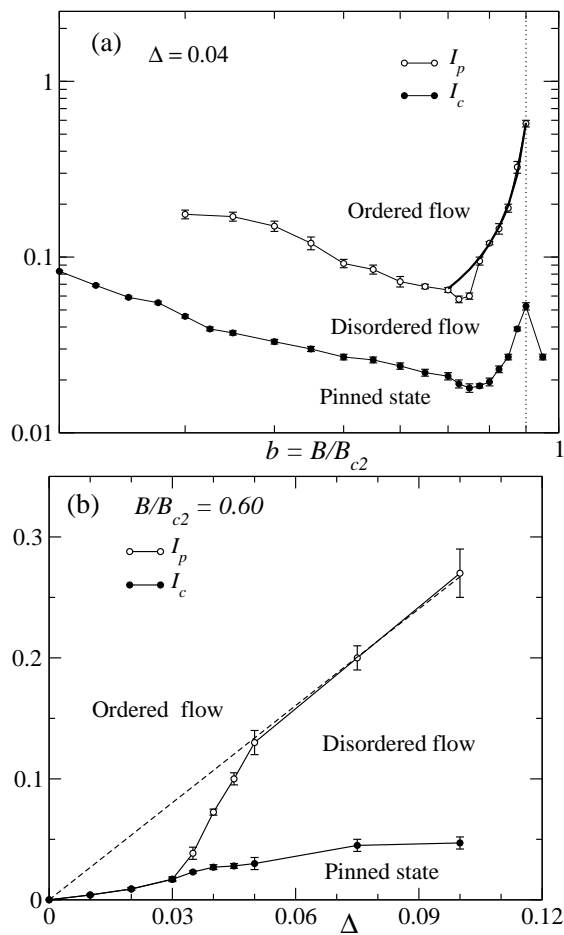
 
\includegraphics[width=210pt]{fig4a}
\includegraphics[width=210pt]{fig4b}
\caption{(a)The dynamic phase diagram showing $I_{p}(b)$ and $I_{c}(b)$ for
$\Delta\!=\!0.04$. The thick line is a fit to $I_{p}(b)\sim(b_{0}-b)^{-1}$ in
the peak effect regime. (b) The $I_p$ and $I_c$ as a function of 
$\Delta$ for $b\!=\!0.60$. The dashed line is a linear fit to the data in the 
range $\Delta\!\geq\!0.05$.}
\end{figure}

Fig.~4(b) shows the $I_{c}$ and the $I_{p}$ as a function of $\Delta$ for
$b\!=\!0.6$. For the system size used in the simulation ($N_{v}\!\sim\!1000$), the
depinning transition is elastic for $\Delta\!\leq\!0.03$ (hence, $I_{p}\!=\!I_{c}$),
whereas for $\Delta\!\geq\!0.05$, $I_{p}\!\propto\!\Delta$. The two disorder regimes
can be identified as the weak pinning and the strong pinning regime, respectively.
The crossover from weak to strong pinning occurs for $\Delta$ between 0.03 and 0.05,
and is also the range in which a sharp dynamic transition is observed. We find that
for the intermediate pinning the real space configuration at $I\!=\!0$ show domains
of ordered lattice separated by domain walls. The presence of ordered region ensures
that the distribution of $v_{x}\!\propto\!\frac{1}{T_{sh}}$ is narrow which
consequently gives a sharp dynamic transition. On the other hand, increasing
$\Delta$ above 0.05 decreases the domain size to $\sim$ 2-3$a_{0}$. This leads to a
broader distribution of $v_{x}$ which in turn broadens the dynamic transition. Thus,
a sharp dynamic transition at $I_p$ implies the existence of ordered regions with
a narrow distribution of the domain size. In Ref.~\cite{paltiel}, the coexistence of 
the ordered and the disordered regions was shown to underlie the shape of the
$V(I)$ curves in the vicinity of the PE. The disordered state in Ref.~\cite{paltiel}
is thought to appear due to the injection of vortices across the surface barrier.
Our simulation suggests that the shape of the $V(I)$ curve depends crucially on the
distribution of the size of the ordered domains in equilibrium.

In conclusion, we have shown that for disordered type-II superconductors, the
dynamic transition current $I_p$ and the critical current $I_c$ shows similar
behaviour as the magnetic field $B$ is varied. $I_p$ decreases with $B$ for fields
below the PE regime. In the peak effect regime, $I_{p}(B)$ increases rapidly along
with the $I_{c}(B)$. The dynamic transition is sharp for the intermediate pinning 
strength due to the presence of large domains of ordered vortex lattice in equilibrium.

M.C. acknowledges useful discussions with E. Zeldov during the course of the work,
and thanks the University of New Mexico for access to their Albuquerque High
Performance Computing Center. This work was supported by NSF-DMR 9985978.


\begin{references}

\bibitem{shi} An-Chang Shi and A. J. Berlinsky, Phys. Rev. Lett. {\bf 67}, 1926 
(1991); H. J. Jensen, A. Brass, and A. J. Berlinsky, Phys. Rev. Lett. {\bf 60}, 1676 
(1988).

\bibitem{kosh} A. E. Koshelev and V. M. Vinokur, Phys. Rev. Lett. {\bf 73}, 3580 
(1994).

\bibitem{giamarchi} T. Giamarchi, and P. Le Doussal, Phys. Rev. Lett. {\bf 76}, 
3408 (1996).

\bibitem{scheidl} S. Scheidl and V. M. Vinokur, Phys. Rev. E {\bf 57}, 2574 (1998).

\bibitem{balents} L. Balents, M. C. Marchetti, and L. Radzihovsky, Phys. Rev. B {\bf 
57}, 7705 (1998).

\bibitem{moon} K. Moon, R. T. Scalettar, and G. T. Zim\'{a}nyi, Phys. Rev. Lett. {\bf 
77}, 2778 (1996).

\bibitem{olson} C. J. Olson, C. Reichhardt, and F. Nori, Phys. Rev. Lett. {\bf 81}, 
3757 (1998).

\bibitem{kolton} A. B. Kolton, D. Dom\'{\i}nguez, and N. Gr$\o$nbech-Jensen, 
Phys. Rev. Lett. {\bf 83}, 3061 (1999).

\bibitem{fangohr} H. Fangohr, S. J. Cox, and P. A. J. de Groot, Phys. Rev. B {\bf 64}, 
064505 (2001).

\bibitem{comment1} At still higher force, the smectic phase goes over to moving Bragg 
glass in 3D for weak disorder~\cite{giamarchi}.

\bibitem{yaron} U. Yaron, P. L. Gammel, D. A. Huse, R. N. Kleimann, C. S. Oglesby, E. 
Bucher, B. Batlogg, D. J. Bishop, K. Mortensen, and K. N. Clausen, Nature {\bf 376}, 753 
(1995).

\bibitem{shobo} S. Bhattacharya and M. J. Higgins, Phys. Rev. Lett. {\bf 70}, 2617
(1993); M. J. Higgins, and S. Bhattacharya, Physica C {\bf 257}, 232 (1996).

\bibitem{duarte} A. Duarte, E. Fernandez Righi, C. A. Bolle, F. de la Cruz, P. L. Gammel, 
C. S. Oglesby, E. Bucher, B. Batlogg, and D. J. Bishop,  Phys. Rev. B {\bf 53}, 11336 
(1996); F. Pardo, F. de la Cruz, P. L. Gammel, E. Bucher, D. J. Bishop, Nature {\bf 396}, 
348 (1998).

\bibitem{chandran1} Mahesh Chandran, R. T. Scalettar, and G. T. Zim\'{a}nyi, (submitted).

\bibitem{chandran2} Mahesh Chandran, cond-mat/0103263.

\bibitem{chandran3} Mahesh Chandran, (unpublished).

\bibitem{paltiel} Y. Paltiel, Y. Myasoedov, E. Zeldov, G. Jung, M. L. Rappaport, D. E. 
Feldman, M. J. Higgins, and S. Bhattacharya, Phys. Rev. B {\bf 66}, 060503(R) (2002).


\end{references}
\end{document}